\documentstyle[preprint,aps]{revtex}
\begin{document}
\draft
\title{Logarithmic Relaxations in a Random Field Lattice Gas \\ 
Subject to Gravity}

\author{Marina Piccioni$^{1}$, Mario Nicodemi$^{1,2}$
        and Serge Galam$^{3}$}
\address{$^1$ Universit\`a di Napoli, Dipartimento di Fisica,
         Mostra d'Oltremare, pad. 19, 80125 Napoli, Italy;
         INFM, Sezione di Napoli}
\address{$^2$INFN, Sezione di Napoli, Napoli, Italy}
\address{$^3$ LMDH, Universit\`e Paris 6, Case 84, Place Jussieu F-75252,
Paris, France}

\maketitle
\begin{abstract}

A simple lattice gas model with random fields and gravity is 
introduced to describe 
a system of grains moving in a disordered environment. 
Off equilibrium relaxations of bulk density and its two time 
correlation functions are numerically found to 
show logarithmic time dependences 
and ``aging" effects. Similitudes with dry granular media are stressed. 
The connections with off equilibrium dynamics in others kinds
of ``frustrated" lattice models in presence of a directional 
driving force (gravity) are discussed to single out the appearance 
of universal features in the relaxation process.

\end{abstract}
%
%
\section{Introduction}

Relaxation properties of granular media in presence of low amplitude
vibrations are dominated by steric hindrance,
friction and inelasticity. They typically show
very long characteristic times \cite{JNBHM}. For instance
the grain compaction process in a box, i.e., the increase of bulk
density in presence of gentle shaking, seems to follow a logarithmic law
\cite{Knight,con,Ben-Naim,DeGennes,tetris}.

This very slow dynamics resemble those of glassy
systems \cite{angell,EvesqueSornette,ConiglioHerrmann}. Moreover
these materials also show ``aging" and ``memory" effects \cite{Knight,NC_hyst}
with logarithmic scaling, and metastability \cite{Knight,NC_hyst}.
The presence of reversible-irreversible cycles (along
which one can identify a sort of ``glassy transition"), as well as
dependence on ``cooling" rates enhance the similarities with glassy systems.
However in granular media, as soon as external forces are cut off, grains
come rapidly to rest. The microscopic origin of
motion is thus quite different from thermal systems (glasses, magnets)
where random particle dynamics
is ensured by temperature. On this basis it is rather surprising to find
those similar behaviors in off equilibrium dynamics.

Based on this resemble, frustrated lattice gas models have been 
recently introduced to
describe these slow logarithmic processes found in granular matter 
(where the shaking amplitude plays the role of
an ``effective" temperature of the system) \cite{con,tetris}.

The aim of the present paper is twofold: on one side 
to describe the correspondences with experiments on granular matter, 
of some dynamical behaviors during 
observed in a frustrated lattice gas model 
subject to gravity; on the other side 
to emphasize the appearance of some very general 
features in the off equilibrium dynamics in presence of gravity 
in a broad class of frustrated lattice gas models.
At first, we thus introduce a very simple random field model to describe
a system of grains moving in a disordered environment.
The relaxation properties of the model are then 
studied via Monte Carlo simulations. We stress the relations 
with granular media and the implications of our results for the  
understanding of the behaviors of such materials. 
Finally, the connections with others kinds
of ``frustrated" models are discussed to single out the universal 
features of this class of lattice systems.

\section{The model}

Along the lines of the Ising Frustrated Lattice Gas (IFLG)
and TETRIS models \cite{con,tetris}, we consider
a system of particles moving on a square lattice
tilted by $45$ degrees.
Because  of the hard core repulsion, every site of the lattice
cannot be occupied by  more than one particle.
Moreover, the system is dilute, so that there
will be empty sites on the lattice.

Although in real systems particles have a rich variety of shapes,
dimensions and orientations in space, here
for the sake of simplicity we restrict ourselves to the case of
one type of grain with an elongated form (like in the TETRIS model
\cite{tetris}) which can lay on the lattice with two different
orientations (Fig.\ref{fig1}).
Each particle is thus characterized by an intrinsic degree of freedom
which is its orientation in space. It can have only
two possible values corresponding to the two possible positions of the
particle on the lattice sites.
Since particles cannot overlap, particles with the same orientation cannot
occupy two nearest neighboring sites on the lattice. As a result a local
geometrical constraint is generated in the dynamics.

In a real granular material each grain moves in a disordered
environment made of the rest of the granular systems itself.
To schematically describe such a situation without entering the full
complexity of the problem, we consider the grains of our model
immersed in a disordered media whose disorder we suppose ``quenched".
This choice may correspond, more strictly, to the case
of grains motion on a geometrically disordered substrate such as a
box with geometric asperities on its surface.
At this stage, the physical feature to embody is the restriction in
the grain motion produced by
the environment disorder.
Therefore, in our model a particle can occupy a lattice site only if its
orientation fits both the local geometry of the medium (Fig.\ref{fig1})
and the geometrical
interaction with the closely neighboring  particles.

These ideas can easily be formalized using
Ising spins in a magnetic language. A spin $S_i$ (with $S_i=\pm 1$) can be
identified with the internal degree of freedom which characterizes the
twofold orientation of particle $i$.
The ``geometrical" interaction between nearest neighboring grains,
which must be antiparallel (to be non-overlapping neighbors),
is realized with antiferromagnetic couplings (of
infinite strength) between nearest neighbors
\cite{nota11}. Analogously,
the random geometry of the environment, which force a grain to have a
definite orientation to fit the local geometry, may be described
as a strong random magnetic field attached to each site of the lattice.

The Hamiltonian of this Random Field Ising system with vacancies,
in presence of gravity, can be thus written as,
\begin{equation}
{\cal H}=J\sum_{\langle{ij}\rangle}(S_{i}S_{j}+1)n_{i}n_{j}-
 H \sum_{i}h_{i}S_{i}n_{i} + g {\sum}_i n_iy_i
\label{eq_Ising}
\end{equation}
where $h_{i}$ are random fields, i.e. quenched variables which assume the
values $\pm 1$, and
$n_{i}$ is the occupancy variable: $n_i=1$ if site $i$ is occupied by
a particle, $n_i=0$ otherwise.
Here $g$ is the gravity constant and $y_i$ corresponds to the height of
the site $i$ with respect to the bottom of the box
(grain mass is set to unity).
$J$ and $H$ represent the repulsions felt by particles if they have a wrong
reciprocal
orientations or if they do not fit the local geometry imposed by the random
fields, respectively. Here we study the case when $J=H= \infty$, i.e.,
the case when the geometric constraints are infinitely
strong. In this case there will be always
sites where ``spins" cannot fulfill
simultaneously every constraint, and thus
will remain vacant.

\section{The tapping}

In a tapping experiment on granular media, a dynamic is imposed to the
grains by vibrations characterized by the shaking normalized intensity
 $\Gamma$ ($\Gamma$ is the ratio of the shake peak acceleration
to the gravity acceleration $g$, see \cite{Knight}).
In the regime of low shaking amplitudes (small $\Gamma$),
the presence of dissipation dominates grain dynamics, and in first
approximation the effects of inertia on particles motion
may be neglected. To embody this scheme in our model we introduce
a two-step diffusive Monte Carlo dynamics for the particles.

The dynamics is very simple with particles moving on the lattice
either upwards or
downwards with respective probabilities $p_{up}$ and
 $p_{down}$ (with $p_{down}=1-p_{up}$).

In the first step, vibrations are on with $p_{up} \ne 0$.
Particles can then
diffuse for a time $\tau_0$ in any direction yet with the inequality
$p_{up}< p_{down}$, and always preserving the above local
geometric
constraints. In the second step,
vibrations are switched
off and the presence of gravity imposes $p_{up}=0$.
Particles can then move only
downwards. In both steps
particle orientation (i.e. its spin $S_i$)
can flip with probability one if there is no
violation of the above constraints and does not flip otherwise.

In this single tap two-step dynamics, we let the system to reach a
{\em static} configuration,
i.e. a configuration in which particles cannot move anymore. In our Monte
Carlo tapping experiment
a sequence of such a vibration is applied to the system.

Under tapping a granular system can move in the space
of available microscopic configurations in a similar way
thermal systems explore their phase space.
The tapping dynamics key parameter is thus the ratio
$x=\frac {p_{up}}{p_{down}}$, which is linked to the experimental
{\em amplitude} of shaking. An effective ``temperature", $T$,
can thus be introduced for the above Hamiltonian with
$x\equiv e^{-\frac{2g}{T}}$ which in turn relates to
granular media real shakes
(see \cite{JNBHM}) via the equality
$\Gamma^a \sim \frac{T}{2g}\equiv 1/\ln(1/x)$, with $a\sim 1,2$
\cite{JNBHM}.

\section{Compaction}

In order to investigate the dynamical properties of the system
when subjected to vibrations, we study the behavior of two
basic observables which are
the density and the density-density time dependent correlation
function. The latter being considered in next section.

The system is initialized filling the container by randomly pouring grains at
the top, one after the other. Particles then fall down subjected only
to gravity, always preserving the model local geometric constraints.
Once they cannot move down any longer, they just stop.
>From this loose packing condition, the system is then shaken by
a sequence of vibrations of amplitude $x=\frac{p_{up}}{p_{down}}$ with
the two steps diffusive dynamics described above.

Specifically, we have studied a $2D$ square lattice of size $30\times 60$
(the results have been observed to be robust to system size changes).
The lattice has periodic
boundary conditions in the horizontal direction and a rigid wall at its
bottom. The particle motion is thus occurring on a cylinder.

After each
vibration $t_n$ (where $t_n$ is the n-th ``tap'')
the bulk density is measured, i.e. the density $\rho (x,t_n)$
in the box lower $25 \%$.

Results for the compaction process are shown in Fig. \ref{fig2}.
Different curves correspond to different values of the amplitude $x$
which ranges from $x=0.001$ up to $x=0.2$.
Data are averaged over $10$ different initial conditions and, for
each initial condition, over $10$ random fields configurations.
The duration of each tap was kept fixed to
$\tau_0=30$ (time is measured in terms of per particle Monte Carlo step)
in all the simulations.
We have also checked that the qualitative general features
exhibited by the model do not depend substantially on the choice of this value.

In analogy to experimental results, the value of $x$ is a crucial parameter
which controls the dynamics of the compaction process and as well as
the final static packing density \cite{Knight}.
In qualitative agreement with experimental findings,
we observe that the stronger is the shaking (i.e., the higher is $x$, or $T$)
the faster the system reaches an higher packing density,
as shown in Fig.\ref{fig2}.
However this is in contrast with the intuitive expectation
about simple systems, as (lattice)
gases in presence of gravity, where higher temperatures correspond to 
lower equilibrium densities \cite{Reif,Stanley}.

It is known experimentally that, by gently shaking a granular system,
the density at the bottom of the box increases very slowly until it reaches
its equilibrium value. The best fit for the density relaxation is an inverse
logarithmic form of the type \cite{Knight},
\begin{equation}
\rho(t_n)={\rho}_{\infty}-\Delta \rho_{\infty}\frac{\log A}{\log
(\frac{t_n}{\tau}+A)}
\label{rho}
\end{equation}
where $\rho_{\infty}$ is the asymptotic density,
$\Delta \rho_{\infty}=\rho_{\infty}-\rho_i$
is the difference between
the asymptotic and initially measured density,
$\rho_i\equiv \rho(t_n=0)\simeq 0.522$,
$A$ and $\tau$ are two fitting  parameters.
Experimentally $\rho_{\infty},
\Delta \rho_{\infty}, A$ and $\tau$ depend on the value of the normalized
shaking amplitude $\Gamma$.

Our different Monte Carlo relaxation curves for the density,
obtained for different intensity of vibration $x$,
show a behavior very similar to the experimental findings and
can be well fitted with the inverse logarithmic law
(\ref{rho}). As stated, the amplitude of shaking $x$ determines the fitting
parameters ($\rho_{\infty}, A$ and $\tau$)
and their behaviors seems to indicate a smooth
crossover between two different regimes at varying $x$.

Fig. \ref{fig3} shows the relaxation of $\tau$ to a constant value with
increasing $x$. The analytical dependence on $x$ can be fitted with a power
law form of the type:
\begin{equation}
\tau \sim \left(B + \frac{1}{x^{\gamma}}\right)
\label{powerlaw}
\end{equation}
$\tau$ can be interpreted as the minimum time over which one starts
to observe a
compaction in the system. For small $x$, $\tau$ exhibits an
algebraic dependence on $x$ with $\gamma= 2.0$, while when $x$ crosses
a certain threshold, say $\tilde{x}\sim 0.1$, it saturates to a constant
value which is independent on the tapping amplitude.

The behavior of the final packing density of the system,
$\rho_{\infty}$, is also shown in Fig.\ref{fig3}.
It increases with increasing $x$ and then, when the shaking intensity
is greater of the typical value $\tilde{x}$, it becomes
almost constant up to the $x$ value we considered.
These results are consistent with those found in Ref.\cite{con},
and are in qualitative agreement with the above described
experimental findings.

It is more difficult to distinguish a definite trend for the value of the
parameter $A$ versus $x$ (see Fig.\ref{fig3}). However it shows a change
of its behavior in
correspondence of a well definite tapping amplitude which has still
approximately the same value $\tilde{x}$.

The recorded behavior of $\rho_{\infty}$ with $x$ (or $\Gamma$ in the
experiments), which, as stated, is in contrast with the expected 
equilibrium values, 
and the observed rough crossover between two regions
suggest that, on typical time scales of both our Monte Carlo
and real experiments, one is typically still far from equilibrium.
Actually if $x$ is sufficiently small, the granular system has
very long characteristic times, as shown by the presence of logarithmic
relaxations. We will address the question of the out of equilibrium
dynamics in next section.

To complete our results about density compaction,
we have plotted in Fig.\ref{fig4} the density profile $\rho (z)$ as
a function of the height from the bottom of the box. Starting from a common
initial configuration, the system is let to evolve while subjected
to shaking with two different
amplitudes $x=0.001$ and $x=0.1$. As already stated, this gives a
difference in the two final density profiles, which seems to approximately 
have (see also \cite{con}) a Fermi-Dirac dependence
on the depth, $z$, as shown in Fig.4: 
\begin{equation}
\rho(z)= \rho_b \left[1-\frac{1}{1+e^{(z-z_0)/s}}\right]
\end{equation}
where $\rho_b$ is the asymptotic bulk density, 
$z_0$ and $s$ are two fitting parameters describing the properties of 
the ``surface'' of the system. They all depend on the shaking
amplitude $x$. The fit parameters for the initial profile are 
$\rho_b=0.53$, $z_0=35.3$ and $s=1.50$; after the shakes with 
$x=0.001$ we find $\rho_b=0.56$, $z_0=36.5$ and $s=0.83$, while 
after the shakes with $x=0.1$ we measure 
$\rho_b=0.575$, $z_0=37.1$ and $s=0.69$. All these show 
that, during compaction, the ``surface'' region of the system shrinks.

\section{The density-density autocorrelation function}

The above discussion about compaction under gentle shaking, with
the presence of logarithmic relaxations 
shows we are in presence of a form of out of equilibrium process.
To characterize its features in a quantitative way, we study
the time dependent correlation functions.

The system is let to evolve for a time interval $t_w$ (the ``waiting time''),
then correlations are measured as a function of time for
$t > t_w$.
In the present case, we record the relaxation features of the two time
density-density correlation function,
\begin{equation}
C(t,t_w)= \frac{\langle \rho (t) \rho (t_w) \rangle - \langle \rho (t) \rangle
\langle \rho (t_w)\rangle}{\langle {\rho (t_w)}^2\rangle -
{\langle \rho (t_w) \rangle}^2}
\label{auto}
\end{equation}
where $<...>$ means the average over a number of random
field configurations and different initial configurations.
As above, $\rho(t)$ is the bulk
density of the system at time $t$.

For a system at equilibrium the time dependent correlation function
$C(t,t_w)$ is invariant under time translations, i.e. it
depends only on the difference $t-t_w$. However,
if the system is off equilibrium,
the subsequent response is
expected to depend explicitly on the
waiting time. Then $C(t,t_w)$ is a function of $t$ and $t_w$ separately.
This ``memory" effect is usually
termed $aging$ and plays an important role in the study
of disordered thermal systems as glassy systems \cite{lp}.

As described above, the initial configuration of the system at $t=0$ is
obtained by pouring grains in a cylindrical box
from the top and letting them
to fall down randomly. For having the system in a well definite configuration
of its parameters,
we started to shake it continuously with a fixed amplitude $x$, i.e.
measures were taken during a single long ``tap".

The results about the bulk density-density correlations of eq.~(\ref{auto}),
averaged over $10$ initial configurations and over $100$ random field
configurations, are shown in Fig.\ref{fig5}.

The various curves correspond to
different
waiting time values, $t_w=72, 360, 720, 7200$ for a fixed shaking amplitude
$x=0.01$.
The dynamics of the system depends strongly on the value of the
waiting time $t_w$, which is the signature of an aging phenomenon.

This behavior is not expected when shaking for long times at high $x$
(i.e., high $T$), where our model is very close to a standard diluted
lattice gas.

The specific scaling of the correlation function describes the system's aging
properties. Therefore, it is interesting to analyze the features
of $C(t,t_w)$ behavior as function of $t$ and $t_w$ for small shaking
amplitudes.

For long enough time,
the correlation function scales with the ratio
$\log (t_w) / \log (t)$. It can thus be approximated
by a scaling form recently introduced to analyze memory effects in IFLG and
TETRIS \cite{NC_hyst},
\begin{equation}
C(t,t_w)=(1-c_{\infty}) \frac{\log(\frac{t_w+t_s}{\tau})}
{\log(\frac{(t+t_s)}{\tau})}+c_{\infty}
\label{corr}
\end{equation}
where $\tau$, $t_s$ and $c_{\infty}$ are fitting parameters.
It is worth noticing these
fitting parameters, for a given $x$, seems to be constant for different
waiting times as shown in (Fig.\ref{fig7}).

All these results on  time dependent correlation
functions thus confirm our picture of an off equilibrium state of the 
dynamics.

\section{Discussion}

The above results thus show logarithmic scaling in off equilibrium 
relaxations of a lattice gas model with antiferromagnetic interactions 
and infinite random fields in presence of gravity. 
These results are, interestingly, consistent with 
the known properties of standard random field Ising systems \cite{bray}. 
Surprisingly, they are also in strong correspondence with 
the behavior of analogous properties observed, by numerical  
investigation, in apparently different lattice models under gravity 
as the quoted TETRIS and IFLG, also introduced to describe granular media
\cite{NC_hyst}. 
The TETRIS is a model which can be mapped, along the same lines outlined 
in the present paper, into a usual lattice gas in presence of 
antiferromagnetic interactions of infinite strength and 
gravity. Its dynamics is ``frustrated" by the presence of purely kinetic 
constraints: particles cannot turn their orientation if too many of their 
neighboring sites are filled.
The IFLG is, instead, a model very close to an Ising Spin Glass 
(also with infinite interaction strengths). It may 
be described in terms of a lattice gas under gravity made of particles 
moving in an environment with quenched disorder. 
The presence of the quoted strong similarities in the off equilibrium 
dynamics of these 
apparently heterogeneous systems suggests the existence of an unexpected 
inherent universality. This seems caused by the important effect of gravity 
on the ``frustrated" dynamics of particles. 
The deep origin of such a phenomenon is yet an open problem and 
it is an important issue to be further investigated.

\section{Conclusions}

In conclusion, we have studied processes of density relaxation in presence
of gentle shaking in a lattice model for granular particles.
The model has a simple geometrical interpretation in terms of
elongated grains moving in a disordered environment and it admits
a mapping into a Random Field Ising system with vacancies, in presence
of gravity. The crucial ingredient in the model is the presence of
geometric frustration dominating particle motion and the necessity
of cooperative rearrangements.
The present model shows logarithmic compaction of its bulk density and
a two time density-density correlation function, $C(t,t')$, which
has an aging behavior well described by a logarithmic scaling
$C(t,t')={\cal{C}}(\ln(t')/\ln(t))$.

At this stage, an experimental check of our finding of aging
in the process of granular media compaction will give a strong ground to
our very simple model.

It is interesting that these results numerically coincide with the 
findings from others two kinds of lattice models which 
seem apparently different:
a model (the TETRIS) which can be mapped into an Ising antiferromagnetic 
lattice gas whose dynamics is characterized by 
purely kinetic constraints, and a model (the IFLG) which is, instead, 
closer to an Ising Spin Glass in presence of gravity. 
The intriguing observation of these similarities seems to suggest the 
presence of a form of universality which appears in ``frustrated" 
particles dynamics, due to the crucial effects of gravity. 

%
%
%
%

%
%
\begin{figure}
\caption{
Schematic picture of the Random Field Frustrated Ising model.
Plus and minus signs represent the orientation of the random field
on each site of the lattice. Filled circles represent the particles 
with the two possible orientations: $S_i=-1$ (white) and $S_i=+1$ (grey).
The coupling between n.n. spins is always antiferromagnetic
(dashed lines). 
Only grey particles can stay on {\em plus} sites and only white
particles can occupy {\em minus} sites.
The result is that each particle on a site of the lattice has to fit
either the geometry imposed but the field either the geometry of the n.n.
spins.
}
\label{fig1}
\end{figure}
\begin{figure}
\caption{
{\rm (a)}
Bulk density (measured in the lower $25 \%$ of the box)
relaxation for tap's amplitude $x=0.001,0.01,0.05,0.1,0.2$
and tap's duration $\tau=30$.
Averages are taken over 10 different initial conditions and for each
initial condition, over 10 random field realizations.
The solid line is a least square fit to the inverse logarithmic form
discussed in the text.}
{\rm (b)}
As in Fig.2a, relaxation of the density measured in the
region on the bottom between $25 \%$ and $50 \%$ in hight of the system.
We find a behavior analogous to what described in Fig.2a.
\label{fig2}
\end{figure}
%
\begin{figure}
\caption{Density relaxation fitting parameter $A, \tau$ and $\rho_{\infty}$
in function of the intensity of vibrations $x$.
In correspondence with experimental findings,
$\rho_{\infty}$ increases with increasing $x$ and, above $\tilde{x}$,
it becomes almost constant.
Such a behavior is in contrast with the intuitive expectation
about simple lattice gases in presence of gravity,
where to higher temperatures should correspond lower equilibrium densities.
}
\label{fig3}
\end{figure}
%
\begin{figure}
\caption{Density profile with increasing the height from the bottom of the box.
Different curves correspond to common initial configuration, and the final
profile for a shaking amplitude $x=0.001$ and $x=0.1$.
Superimposed are the Fermi-Dirac fits described in the text.} 
\label{fig4}
\end{figure}
%

\begin{figure}
\caption{The two time density-density correlation function, $C(t,t_w)$,
for a fixed vibrations amplitude $x=10^{-2}$.
The different curves correspond to four ``waiting time''
$t_w=72, 360,720,7200$. $C(t,t_w)$ is not a simple function of
$t-t_w$ and shows ``aging''.
}
\label{fig5}
\end{figure}

%
\begin{figure}
\caption{The fitting parameters from eq.~(6)
of the correlation function, $C(t,t_w)$, at $x=10^{-2}$,
seems to be almost independent of the waiting time $t_w$.}
\label{fig7}
\end{figure}

\begin{references}
\bibitem{nota11} This part of the mapping Hamiltonian may depend on
the details, as shapes and sizes, of the grains taken into account.
We consider here the simplest non trivial case.
\end{references}
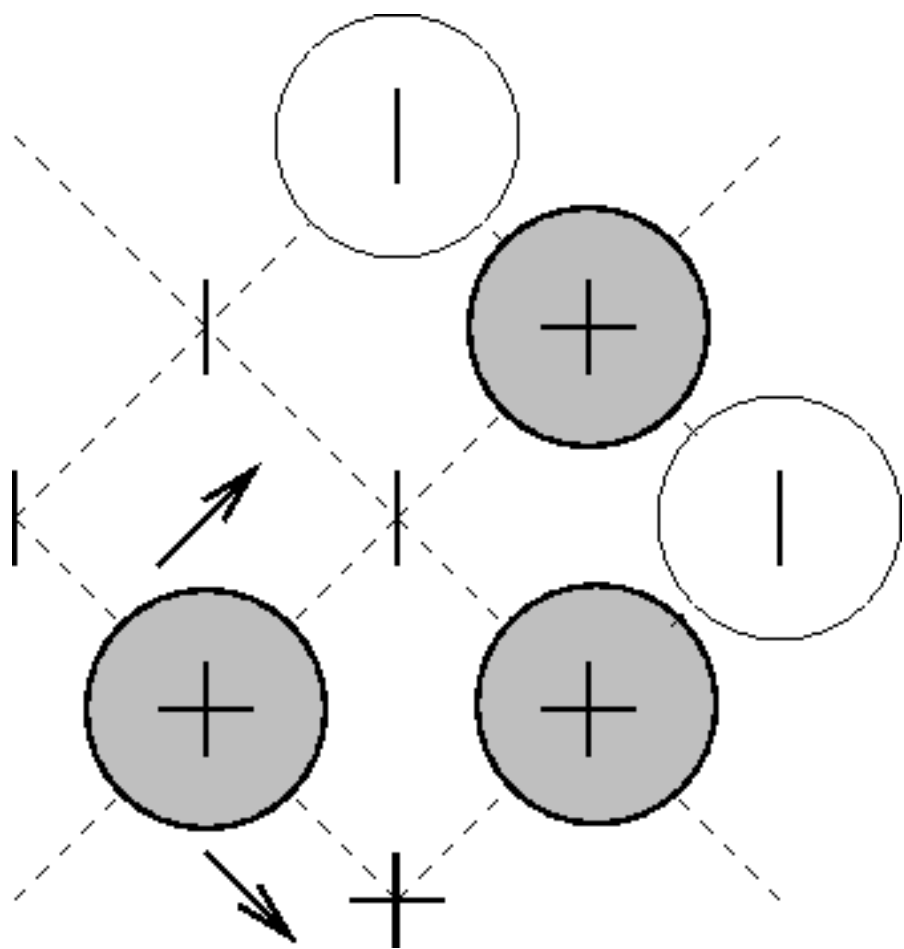
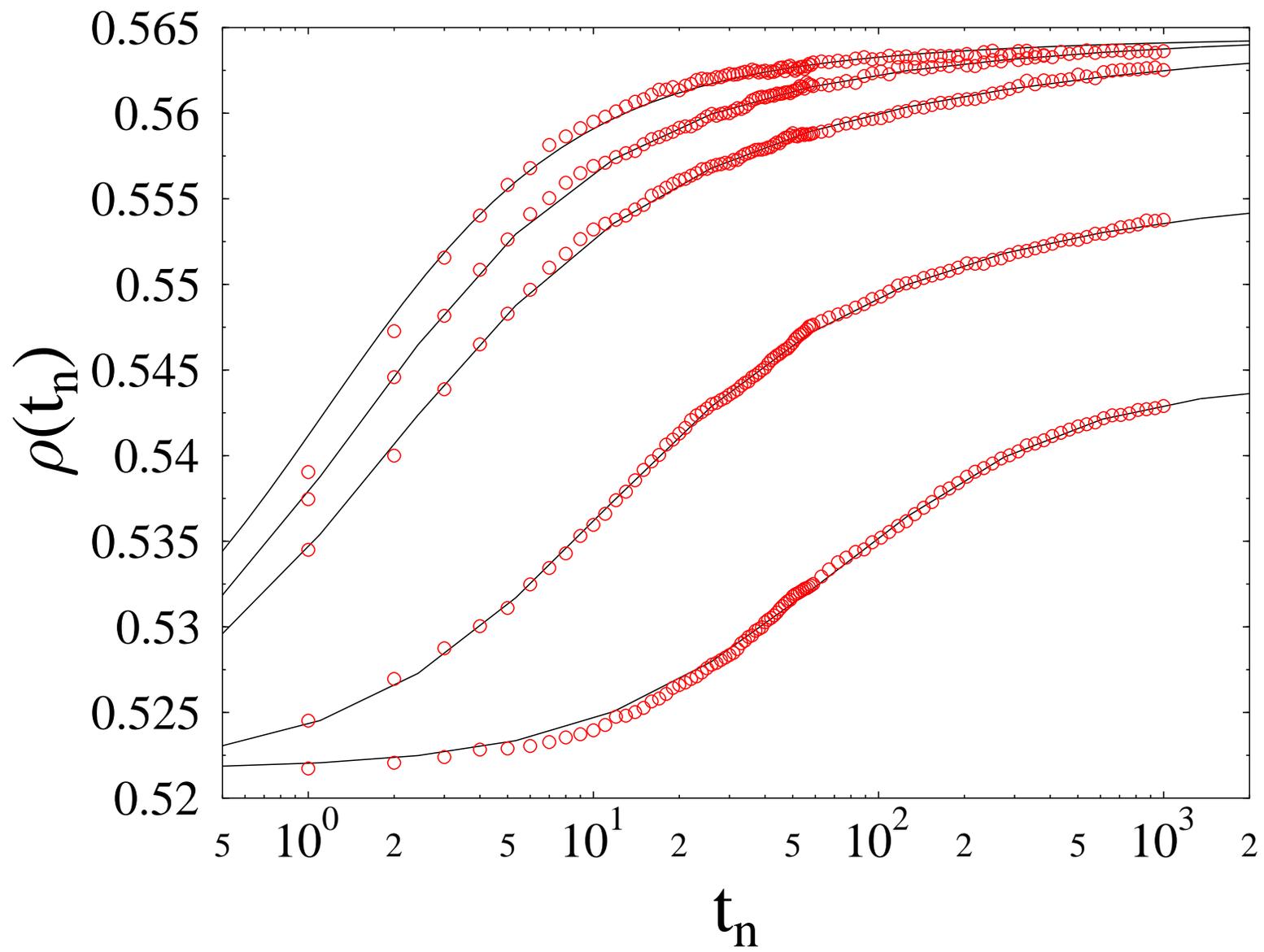
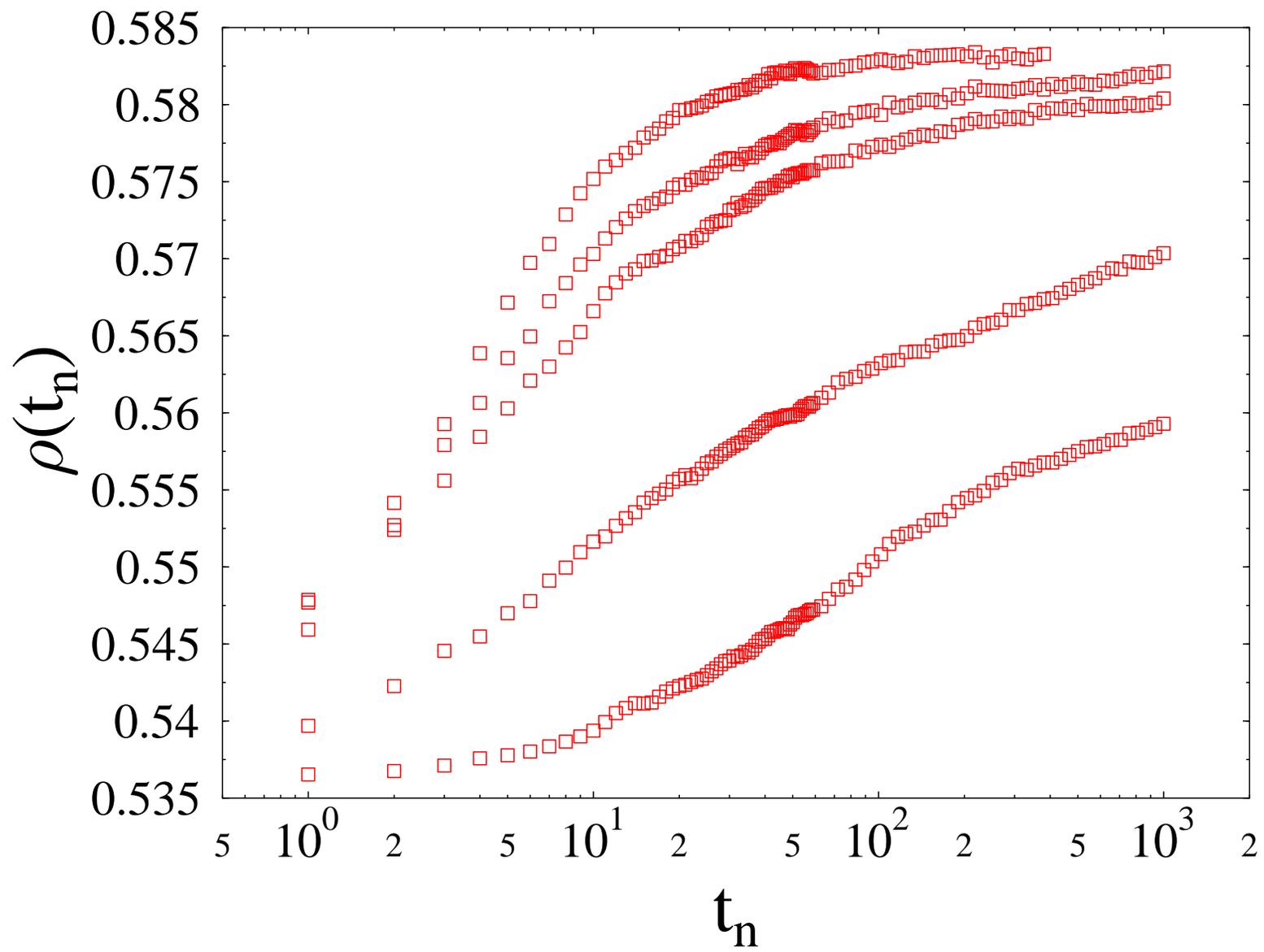
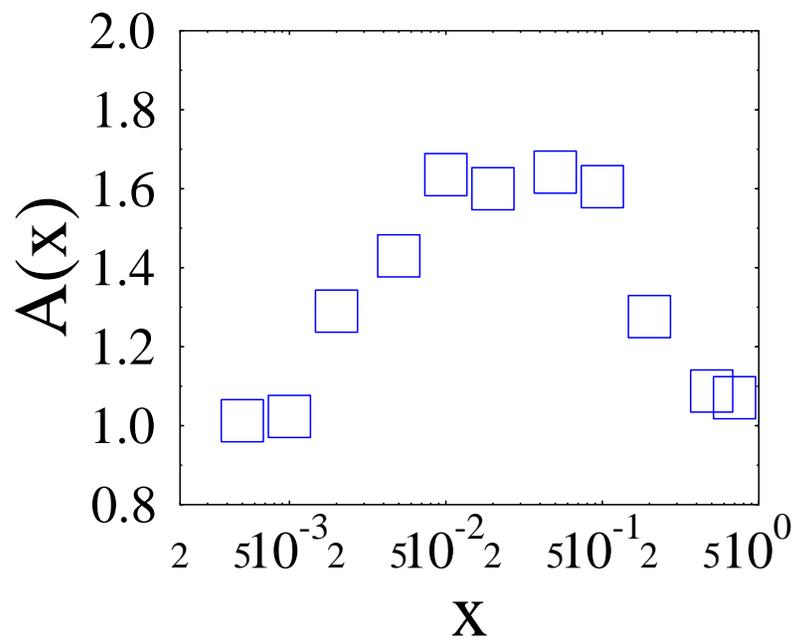
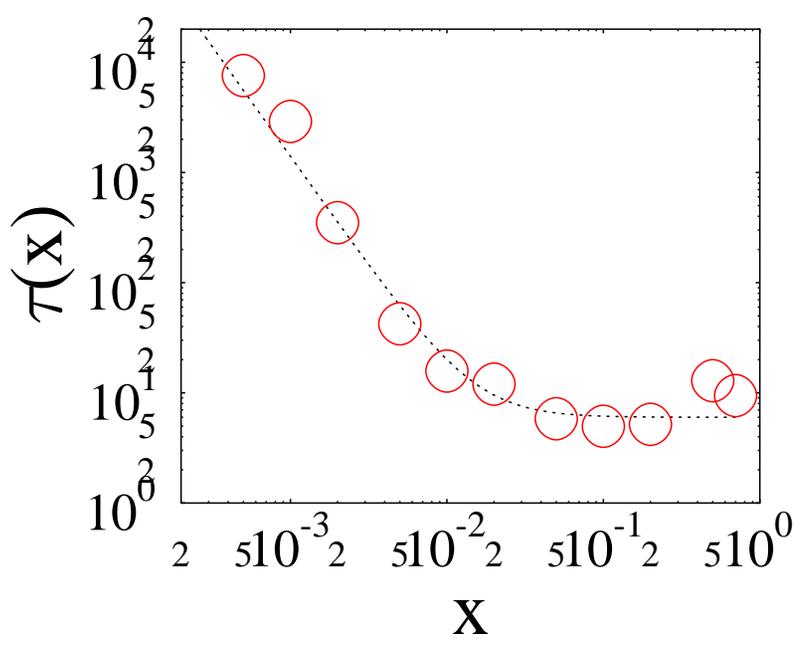
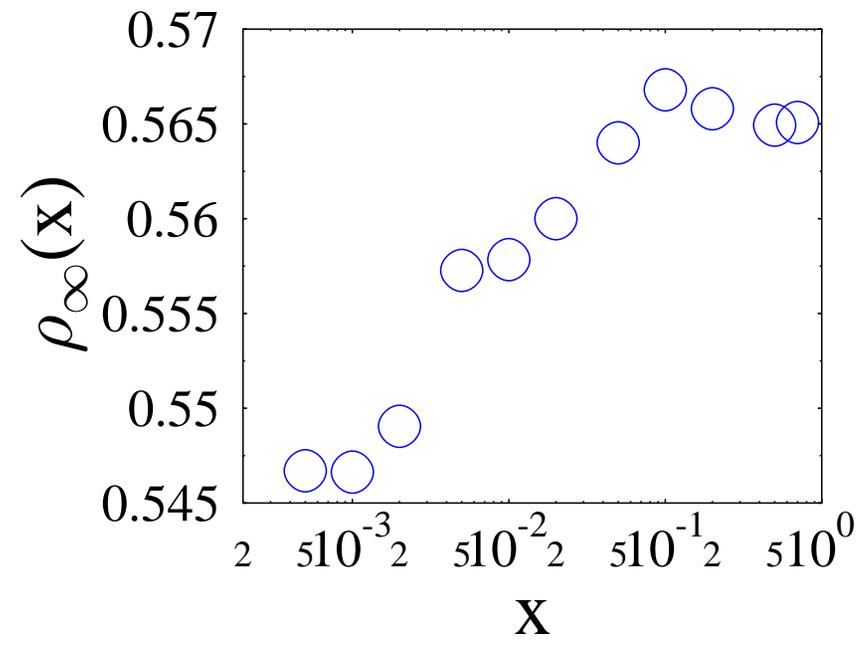
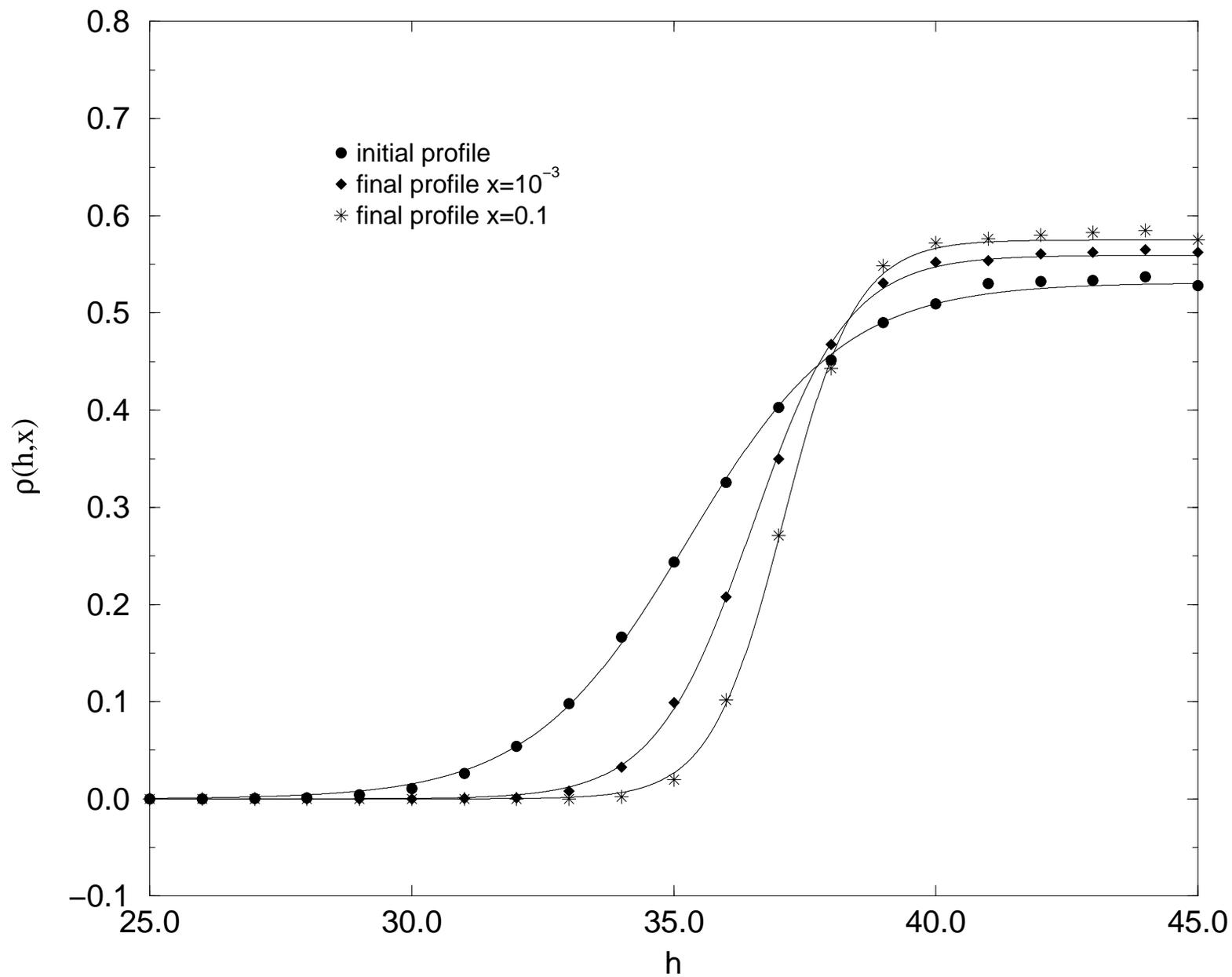
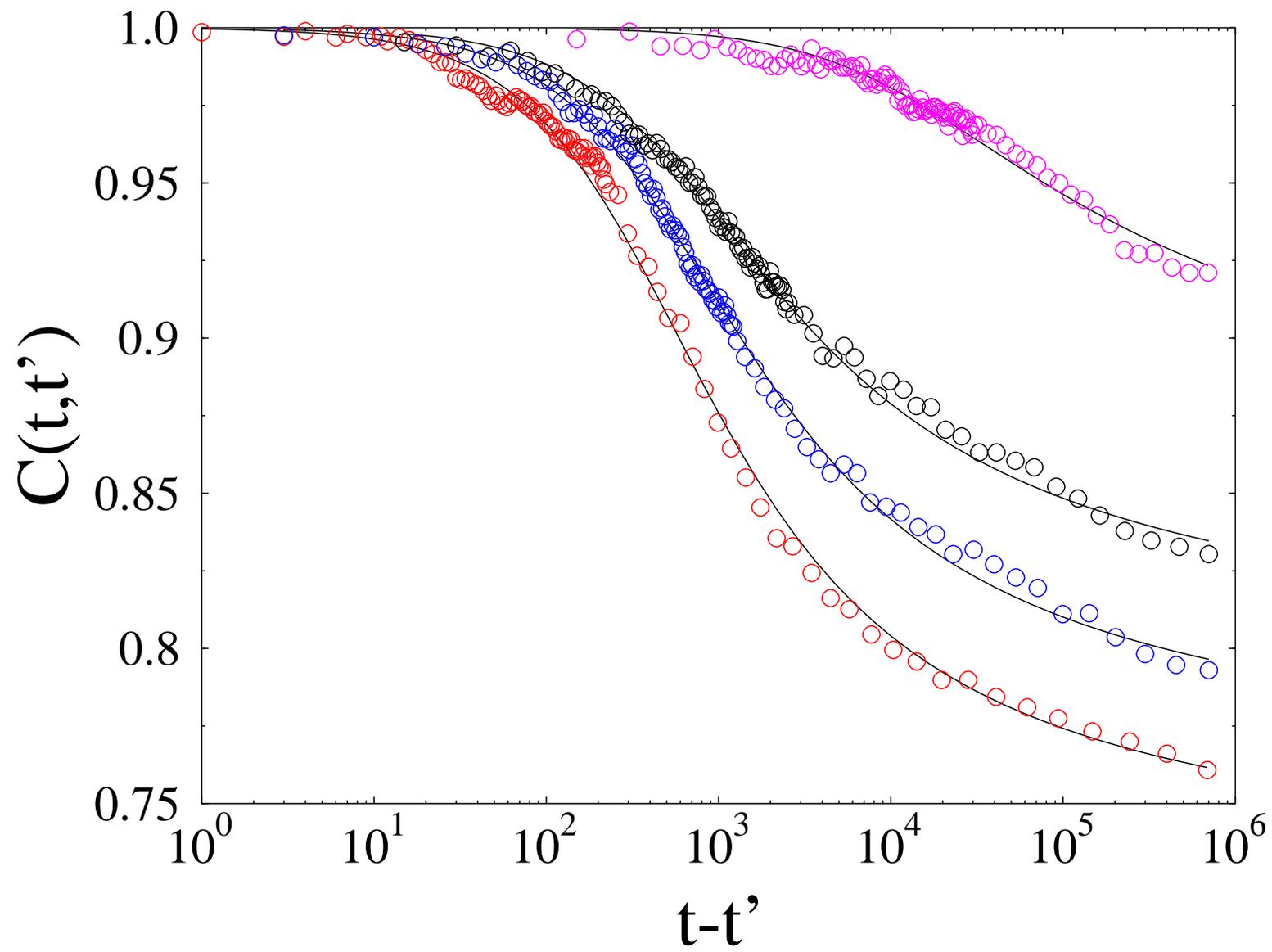
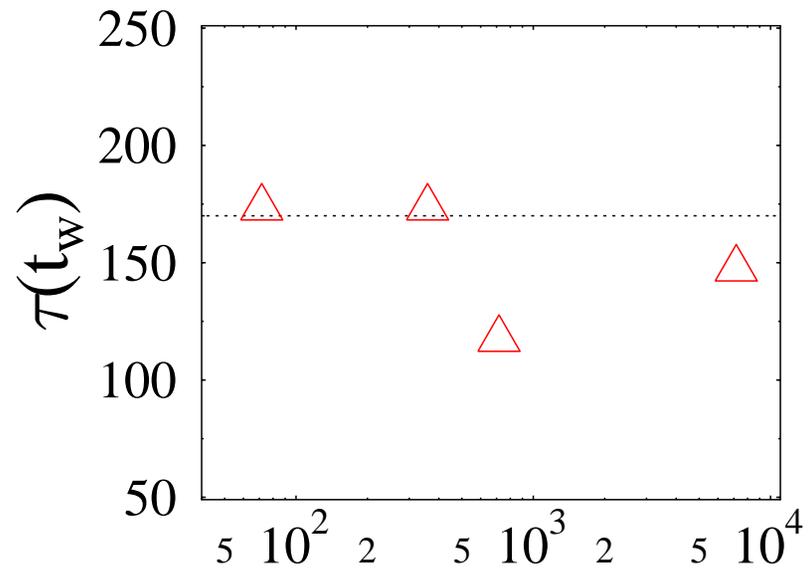
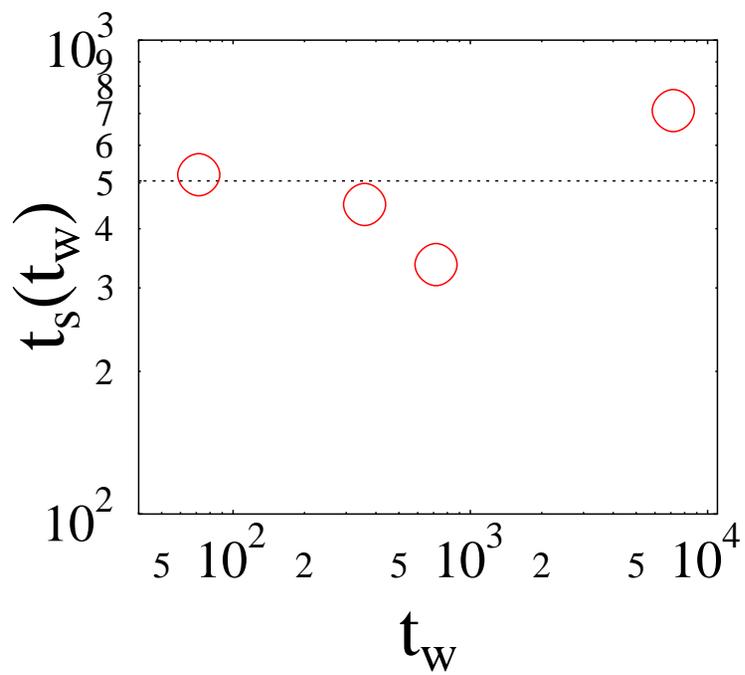
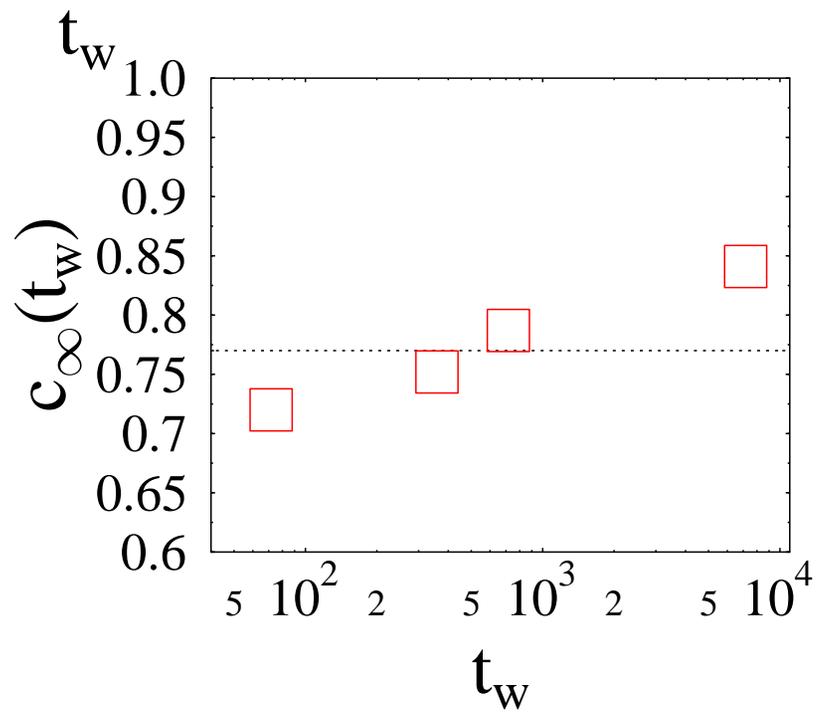
\end{document}